\documentclass[showpacs,aps,twocolumn]{revtex4}
\usepackage{graphicx}
\usepackage{amsfonts}
\usepackage{amsmath}
\usepackage{amssymb}
\usepackage{multirow}
\usepackage{braket}
\usepackage[usenames,dvipsnames]{color}

\begin{document}
\bibliographystyle{apsrev}

\title{Entropy of finite random binary sequences with weak long-range
correlations}

\author{ S.~S.~Melnik\footnote[1]{melnikserg@yandex.ru} and O.~V.~Usatenko
\footnote[2]{usatenko@ire.kharkov.ua}}
\affiliation{ A. Ya. Usikov Institute for Radiophysics and Electronics \\
Ukrainian Academy of Science, 12 Proskura Street, 61805 Kharkov,
Ukraine}

\begin{abstract}

We study the $N$-step binary stationary ergodic Markov chain and
analyze its differential entropy. Supposing that the correlations
are weak we express the conditional probability function of the
chain through the pair correlation function and represent the
entropy as a functional of the pair correlator. Since the model uses
the two-point correlators instead of the block probability, it makes
it possible to calculate the entropy of strings at much longer
distances than using standard methods. A fluctuation contribution to
the entropy due to finiteness of random chains is examined. This
contribution can be of the same order as its regular part even at
the relatively short lengths of subsequences. A self-similar
structure of entropy with respect to the decimation transformations
is revealed for some specific forms of the pair correlation
function. Application of the theory to the DNA sequence of the R3
chromosome of \emph{Drosophila melanogaster} is presented.
\end{abstract}

 \maketitle


\section{Introduction}

At present there is a commonly accepted viewpoint that our world is
complex and correlated. The most peculiar manifestations of this
concept are the records of brain activity and heart beats, human and
animal communication, written texts of natural languages, DNA and
protein sequences, data flows in computer networks, stock indexes,
sun activity, weather (the chaotic nature of the atmosphere), etc.
For this reason systems with long-range interactions (and/or with
long-range memory) and natural sequences with non-trivial
information content have been the focus of a large number of studies
in different fields of science over the past several decades.

Random sequences with finite number of states exist as natural
sequences (DNA or natural texts) or arise as a result of
coarse-grained mapping of the evolution of the chaotic dynamic
system into a string of symbols~\cite{Eren,Lind}. Such sequences are
very closely connected to and are the subject of study of the
algorithmic (Kolmogorov-Solomonoff-Chaitin) complexity, artificial
intellect, information theory, compressibility of digital data,
statistical inference problem, computability~\cite{Cover} and have
many application aspects as a creative tool for designing the
devices and appliances with random components in their
structure~\cite{IzrKrMak} (different wave-filters, diffraction
gratings, artificial materials, antennas, converters, delay lines,
etc.).

There are many methods for describing complex dynamical systems and
random sequences connected with them: correlation function, fractal
dimensions, multi-point probability distribution functions, and many
others. One of the most convenient characteristics serving to the
purpose of studying complex dynamics is entropy~\cite{Shan,Cover}.
Being a measure of the information content and redundancy in a
sequence of data, it is a powerful and popular tool in examination
of complexity phenomena. It has been used for the analysis of a
number of different dynamical systems.

A standard method of understanding and describing statistical
properties of real physical systems or random sequences of data can
be represented as follows. First of all, we have to analyze the
sequence to find the correlation functions or the probabilities of
words occurring, with the length $L$ exceeding the correlation
length $R_{c}$ but being shorter than the length $M$ of the
sequence,
\begin{equation} \label{Strong}
 R_{c}<L\ll M.
\end{equation}
At the same time, the number $d^L$ of different words of the length
$L$ composed in the alphabet containing $d$ letters has to be much
less than the number $M-L$ of words in the sequence,
\begin{equation}\label{Strong2}
 d^L\ll M-L\approx M.
\end{equation}
The next step is to express the correlation properties of the
sequence in terms of the conditional probability function (CPF) of
the Markov chain, see below Eq.~(\ref{soglas}). Note, the Markov
chain should be of order $N$, which is supposed to be longer than
the correlation length,
\begin{equation}\label{Strong3}
R_{c}<N.
\end{equation}
This is the critical requirement because the correlation length of
natural sequence of interest (e.g., written or DNA texts) is usually
of the same order as the length of sequences. None of inequalities
$(\ref{Strong})-(\ref{Strong3})$ can be fulfilled. Really, the
lengths of words that could represent correctly the probability of
words occurring are $4-5$ letters for a real natural text of the
length $10^6$ (written on an alphabet containing $27-30$ letters and
symbols) or of order of 20 symbols for a coarse-grained text
represented by means of a binary sequence.

Here we develop an approach that is complimentary to the above
exposed. In particular, we represent the conditional probability
function of the Markov chain by means of pair correlator, which
makes it possible to calculate analytically the entropy of the
sequence. It should be stressed that the standard method for
calculating the entropy can only take into account the short-range
part of statistics. We present a theory that expresses long-range
correlation properties through the correlation functions.

The scope of the paper is as follows. First, we discuss briefly the
properties of the $N$-step additive Markov chain model and,
supposing that the correlations between symbols in the sequence are
weak, we express the conditional probability function by means of
the pair correlation function. In the next section we represent the
differential entropy in terms of the conditional probability
function of the Markov chain and express the entropy as the sum of
squares of the pair correlators. Then we discuss some properties of
the results obtained, in particular, a property of self-similarity
of entropy with respect to decimation for some particular classes of
the Markov chains. Next, a fluctuation contribution to the entropy
due to finiteness of random chains is examined. Some remarks on
literary texts are followed by discussions of directions in which
the research can be progressed.

\section{Additive Markov chains}

This section includes mainly introductory material. Some authors'
results presented by Eqs.~(\ref{prob_series})--(\ref{Approx CP})
were exposed earlier in Ref.~\cite{muya,RewUAMM,MUYaG,highord,UYa}.

Consider a semi-infinite sequence $\mathbb{A}=a_{0}^{\infty}= a_{0},
a_{1},a_{2},...$ of real random variables $a_{i}$ taken from the
finite alphabet $A=\{1,2,...,d\}$, $a_{i}\in A$. The sequence
$\mathbb{A}$ is an \textit{$N$-step Markov chain} if it possesses
the following property: the probability of symbol~$a_i$ to have a
certain value $a$ under the condition that the values of all other
symbols are given depends only on the values of $N$ previous
symbols,
%
%
\begin{eqnarray}\label{def_mark}
&& P(a_i=a|\ldots,a_{i-2},a_{i-1})\nonumber\\[6pt]
&&=P(a_i=a|a_{i-N},\ldots,a_{i-2},a_{i-1}).
\end{eqnarray}

Note, definition~(\ref{def_mark}) is valid for $i\geq N$; for $i<N$
we have to use the well known conditions of compatibility for the
conditional probability functions of the lower order,
Ref.~\cite{shir}. Sometimes the number $N$ is also referred  to as
the \emph{order} or the \emph{memory length} of the Markov chain.
The conditional probability function (CPF)
$P(a_i=a|a_{i-N},\ldots,a_{i-2},a_{i-1})$ determines completely all
statistical properties of the Markov chain and the method of their
iterative numerical construction. If the sequence,  statistical
properties of which we would like to analyze is assigned, the
conditional probability function of the $N$-th order can be found by
a standard method,
\begin{equation}\label{soglas}
P(a_{N+1}=a|a_{1},\ldots,a_{N})=\frac{ P(a_{1},\ldots,a_{N},a) } {
P(a_{1},\ldots,a_{N})},
\end{equation}
where $P(a_{1},\ldots,a_{N})$ is the probability  of the $N$-words
$a_{1},\ldots,a_{N}$ occurring.

The Markov chain determined by Eq.~(\ref{def_mark}) is a
\textit{homogeneous} sequence because its conditional probability
does not depend explicitly on $i$, i.e., is independent of the
position of symbols $a_{i-N},\ldots,a_{i-1},a_{i}$ in the chain. It
depends only on the values of symbols
$a_{i-N},\ldots,a_{i-1},a_{i}$. The homogeneous sequences are
\emph{stationary}: the average value of any function
$f(a_{r_1},a_{r_1+r_2},\ldots , a_{r_1+\ldots+r_{s}})$  of $s$
arguments
\begin{eqnarray}\label{epsilon-av}
&& \overline{f}\,(a_{r_1},\ldots ,
a_{r_1+\ldots+r_{s}})\\[6pt]
&&=\lim_{M\to\infty}\frac{1}{M} \sum_{i=0}^{M-1}f(a_{i+r_1},\ldots ,
a_{i+r_1+\ldots+r_{s}}) \nonumber
\end{eqnarray}
depends on $s-1$ differences between the indexes. In other words,
all statistically averaged functions of random variables are
\emph{shift-invariant}.

We suppose that the chain is \emph{ergodic}. According to the Markov
theorem (see, e.g., Ref.~\cite{shir}), this property is valid for
the homogenous Markov chains if the strict inequalities,
\begin{equation}\label{ergo_m}
0 < P(a_{i+N}=\alpha|a_{i}^{i+N-1}) < 1, \, i \in \mathbb{N}_{+} =
\{0,1,...\}
\end{equation}
are fulfilled for all possible values of the arguments in function
(\ref{def_mark}). Hereafter we use the shorter notation
$a_{i-N}^{i-1}$ for $N$-word $a_{i-N},...,a_{i-1}$. It follows
from ergodicity that the correlations between any blocks of symbols
in the chain go to zero when the distance between them goes to
infinity. The other consequence of ergodicity is the possibility to
use one random sequence as an equitable representative of the
ensemble of chains and to do averaging over the sequence,
Eq.~(\ref{epsilon-av}), instead of an ensemble averaging.

Below we will consider an important class of the binary random
sequences with symbols~$a_i$ taking on two values, say $0$~and~$1$,
$a_i \in \{0, 1\}$. The conditional probability to find $i$-th
element $a_i=1$ in  the \emph{binary} $N$-step Markov sequence
depending on $N$ preceding elements $a_{i-N}^{i-1}$ is a set of
$2^N$ numbers:
\begin{eqnarray} \label{prob}
&& P(1|a_{i-N}^{i-1})=P(a_{i}=1|a_{i-N}^{i-1}),\nonumber\\[6pt]
&&P(0|a_{i-N}^{i-1}) =1- P(1|a_{i-N}^{i-1}).
\end{eqnarray}
Conditional probability (\ref{prob}) of the binary sequence of
random variables $a_i\in \{0, 1\}$ can be represented exactly as a
\emph{finite} polynomial series:
\begin{eqnarray} \label{prob_series}
&&P(1|a_{i-N}^{i-1}) = \bar{a} + \sum_{r_1=1}^N F_1(r_1)(a_{i-r_1}
- \bar{a}) \nonumber\\
    &&+\sum_{r_1,r_2=1}^N F_2(r_1,r_2)(a_{i-r_1} a_{i-r_2} -
    \overline{a_{i-r_1} a_{i-r_2}}) +
    \ldots  \nonumber\\
    &&+\sum_{r_1,\ldots,r_N=1}^N F_N(r_1,\ldots,r_N)(a_{i-r_1} \ldots
    a_{i-r_N}\nonumber\\
    &&- \,\,\overline{a_{i-r_1} \ldots
    a_{i-r_N}}),
\end{eqnarray}
where the statistical averages $\overline{a_{r_1} \ldots
    a_{r_N}}$ are taken over
sequence~(\ref{epsilon-av}), $F_s$ is the family of \textit{memory
functions} and $\bar{a}$ is the relative average number of unities
in the sequence. The representation of Eq.~(\ref{prob}) in the form
of Eq.~(\ref{prob_series}) follows from  the simple identical
equalities, $a^2=a$ and $f(a) = a f(1) + (1-a) f(0)$,  for an
arbitrary function $f(a)$ determined on the set $a\in \{0, 1\}$. The
first term in Eq.~(\ref{prob_series}) is responsible for generation
of uncorrelated white-noise sequences. Taking into account the
second term, proportional to $F_1(r)$, we can reproduce correctly
correlation properties of the chain up to the second order.
Higher-order correlators and all correlation properties of higher
orders are not independent anymore. We cannot control them and
reproduce correctly by means of the memory function $F(r)$, because
the latter is completely determined by the pair correlation
function, see below Eq.~(\ref{main}). Studying of the properties of
these higher-order correlators is beyond the scope
of this paper. 
In what follows we will only
use the first two terms, which determine the so-called
\emph{additive} Markov chain~\cite{muya,RewUAMM}.

A particular form of the conditional probability function of
additive Markov chain is the chain with step-wise memory function,
\begin{eqnarray} \label{prob step}
P(1|k) = \frac{1}{2} + \mu \left(\frac{2k}{N} -1\right).
\end{eqnarray}
The probability $P(1|k)$ of having the symbol $a_i=1$ after $N$-word
$a_{i-N}^{i-1}$ containing $k$ unities, $\,\, k=\sum_{l=1}^{N}
a_{i-l}$, is a linear function of $k$ and is independent of the
arrangement of symbols in the word $ a_{i-N}^{i-1}$. The parameter
$\mu$ characterizes the strength of correlations in the system.

There is a rather simple relation between the memory  function
$F(r)$ (hereafter we will omit the subscript $1$ of $F_1(r)$) and
the pair correlation function of the binary additive Markov chain.
There were suggested two methods for finding the $F(r)$ of a
sequence with a known pair correlation function. The first
one~\cite{muya} is based on the minimization of a ``distance''\,
between the Markov chain generated by means of the sought-for memory
function and the initial given sequence of symbols with a known
correlation function. The minimization equation yields the
relationship between the correlation and memory functions,
\begin{equation} \label{main}
K(r)=\sum\limits_{r'=1}^{N}F(r')K(r-r'), \ \ \ \ r\geqslant 1.
\end{equation}
where the normalized correlation function $K(r)$ is given by
\begin{equation}\label{def_cor1}
K(r)=\frac{C(r)}{C(0)}, \quad
C(r)=\overline{(a_i-\bar{a})(a_{i+r}-\bar{a})}.
\end{equation}
The second method for deriving Eq.~(\ref{main}) is the completely
probabilistic straightforward calculation~\cite{MUYaG}.

Equation~(\ref{main}), despite its simplicity,  can be analytically
solved only in some particular cases: for one- or two-step chains,
Markov chain with step-wise memory function and so on. To avoid the
difficulties in solving Eq.~(\ref{main}) we suppose that
correlations in the sequence are weak. This means that all
components of the normalized correlation function are small,
$|K(r)|\ll 1, \,|r|\neq 0$, with the exception of $K(0)= 1$. So,
taking into account that in the sum of Eq.~(\ref{main}) the leading
term is $K(0)=1$ and all the others are small, we can obtain an
approximate solution for the memory function in the form of the
series
\begin{eqnarray} \label{Series}
F(r)&=&K(r) - \sum_{r'\neq r}^N K(r-r') K(r') \\
&+&\sum_{r'\neq r}^N\sum_{r''\neq r'}^N K(r-r')
K(r'-r'')K(r'')+...\nonumber
\end{eqnarray}
The equation for the conditional probability function in the first
approximation with respect to the small functions $|K(r)|\ll 1,
\,|r|\neq 0$ takes the form
\begin{eqnarray} \label{Approx CP}
P(1|a_{i-N}^{i-1})&\simeq &\bar{a} + \sum_{r=1}^N F(r)(a_{i-r} -
\bar{a}) \nonumber\\
&\simeq &\bar{a} + \sum_{r=1}^N K(r)(a_{i-r} - \bar{a}).
\end{eqnarray}
This formula provides a very important tool for constructing a
sequence with a given pair correlation function. Note that
$i$-independence of the function $P(1|a_{i-N}^{i-1})$ guarantees
homogeneity and stationarity of the sequence under consideration;
and finiteness of $N$ provides its ergodicity. Evidently, we can
only consider sequences with the correlation functions, determined
by $P(1|a_{i-N}^{i-1})$, which satisfy Eq.~(\ref{ergo_m}).

The correlation functions are typically employed as the input
characteristics for describing the random sequences. However, the
correlation function describes not only the direct interconnection
of the elements $a_i$ and $a_{i+r}$, but also takes into account
their indirect interaction via all other intermediate elements. Our
approach operates with the ``origin'' characteristics of the system,
specifically, with the memory function. The correlation and memory
functions are mutually complementary characteristics of a random
sequence in the following sense. The numerical analysis of a given
random sequence enables one to determine directly the correlation
function rather than the memory function. On the other hand, it is
possible to construct a random sequence using the memory function,
but not the correlation one, in the general case. Therefore, the
memory function permits one to get a deeper insight into the
intrinsic properties of the correlated systems.
Equation~(\ref{Approx CP}) shows that in the limit of weak
correlations both functions play the same role.

The concept of the additive Markov chain was extensively used
earlier for studying random sequences with long-range correlations.
The examples and references can be found in Ref.~\cite{RewUAMM}.

\section{Differential entropy}

In order to estimate the entropy of an infinite stationary sequence
$\mathbb{A}$ of symbols $a_{i}$ one could use the block entropy,
\begin{eqnarray} \label{entro_block}
H_{L}=-\sum_{a_{1},...,a_{L}} P(a_{1}^{L})\log_{2}
P(a_{1}^{L}).
\end{eqnarray}
Here $P(a_{1}^{L}) =P(a_{1},\ldots,a_{L})$ is the probability to
find the $L$-word $a_{1}^{L}$ in the sequence. The differential
entropy, or entropy per symbol, is given by
\begin{eqnarray} \label{entro_diff}
h_{L}=H_{L+1} - H_{L},
\end{eqnarray}
and specifies the degree of uncertainty of the $(L+1)$th symbols
observing if the preceding $L$ symbols are known. The source entropy
(or Shannon entropy) is the differential entropy at the asymptotic
limit, $h=\lim_{L \rightarrow \infty}h_{L}$. This quantity  measures
the average information per symbol if {\it all} correlations, in the
statistical sense, are taken into account.

The differential entropy $h_L$ can be presented in terms of the
conditional probability function. To show this we have to rewrite
Eq.~\eqref{entro_block} for the block of length $L+1$,  express
$P(a_{1}^{L+1})$ via the conditional probability, and after a bit of
algebra we obtain 
\begin{eqnarray} \label{Entro_Bin}
h_L=\!\!\!\sum_{a_{1},...,a_{L}=0,1} \!\!\!P(a_{1}^{L})
h(a_{L+1}|a_{1}^{L}) = \overline{ h(a_{L+1}|a_{1}^{L})}.
\end{eqnarray}
Here $h(a_{L+1}|a_{1}^{L})$ is the amount of information contained
in the $(L+1)$-th symbol of the sequence conditioned on $L$ previous
symbols,
\begin{eqnarray}
   h(a_{L+1}|a_{1}^{L}) = - \!\!\!\sum_{a_{L+1}=0,1}\!\!\!
P(a_{L+1}|a_{1}^{L})\log_2 P(a_{L+1}|a_{1}^{L}).
    \label{siL}
\end{eqnarray}
So, the differential entropy $h_L$ of random sequence is presented
as a special case of the standard conditional entropy $H=-\sum_{C}
P(C) \sum_{B} P(B|C)\log_2 P(B|C)$.

The conditional probability $P(1|a_{i-L}^{i-1})$ at $L < N$,
\begin{eqnarray} \label{p_i(L)}
P(1|a_{i-L}^{i-1}) \simeq \bar{a} + \delta; \quad \delta =
\sum_{r=1}^L F(r)(a_{i-r} -\bar{a}),
\end{eqnarray}
is obtained in the first approximation in the parameter $\delta$
from Eq.~\eqref{Approx CP} by means of the probabilistic reasoning
presented in the Appendix.

Taking into account the weakness of correlations, $|\delta| \ll \min
[\overline{a}, (1-\overline{a})]$, one can expand the right-hand
side of Eq.~\eqref{siL} in Taylor series up to the second order in
$\delta$, $h(a_{L+1}|a_{1}^{L}) = h_0 + (\partial
h/\partial\overline{a})_{|\delta=0}\delta + (1/2)(\partial^2
h/\partial\overline{a}^2)_{|\delta=0}\delta^2$, where the
derivatives are taken at the ``equilibrium point''
$P(1|a_{i-L}^{i-1}) = \bar{a}$ and $h_0$ is the entropy of
uncorrelated sequence,
\begin{eqnarray}
h_0=-\bar{a}\log_{2}(\bar{a}) - (1-\bar{a})\log_{2}(1-\bar{a}).
\end{eqnarray}
Upon using Eq.~\eqref{Entro_Bin} for averaging
$h(a_{L+1}|a_{1}^{L})$ and in view of $ \overline{\delta} =0$, the
differential entropy of the sequence becomes
\begin{equation}\label{Entro_Markov2}
h_L= \left\{\begin{array}{l} h_{L \leq N}= h_0 -
\dfrac{1}{2\ln2}\sum_{r=1}^L F^2(r), \\
 h_{L>N}=h_{L=N}.
\end{array}
\right.
\end{equation}
If the length of block exceeds the memory length, $L>N$, the
conditional probability $P(1|a_{i-L}^{i-1})$ depends only on $N$
previous symbols, see Eq.~(\ref{def_mark}). Then, it is easy to show
from~\eqref{Entro_Bin} that the differential entropy remains
constant at $ L \ge N$. The second line of Eq.~\eqref{Entro_Markov2}
is consistent  with the first one because in the first approximation
in $\delta$ the correlation function vanishes at $L>N$ together with
the memory function.  The final expression, the main result of the
paper,  for the differential entropy of the stationary ergodic
binary weakly correlated random sequence is
\begin{equation}\label{EntroMain}
h_L=  h_0 - \frac{1}{2\ln2}\sum_{r=1}^L K^2(r).
\end{equation}

\section{Discussion}

It follows from Eq.~(\ref{EntroMain}) that the additional correction
to the entropy $h_0$ of the uncorrelated sequence is the negative
and monotonously decreasing function of $L$. This is the anticipated
result --- the correlations decrease the entropy. The conclusion is
not sensitive to the sign of correlations: persistent correlations,
$K>0$, describing an ``attraction'' of the symbols of the same kind,
and anti-persistent correlations, $K<0$, corresponding to an
attraction between ``0'' and ``1'', provide the corrections of the
same negative sign. If the correlation function is constant at $1
\leqslant r \leqslant N$, the entropy is a linear decreasing
function of the argument $L$ up to the point $N$; the result is
coincident with that obtained in Ref.~\cite{DMBUYa} (in the limit of
weak correlations) for the Markov chain model with step-wise memory
function~(\ref{prob step}).

\begin{figure}[h!]
\includegraphics[width=0.5\textwidth]{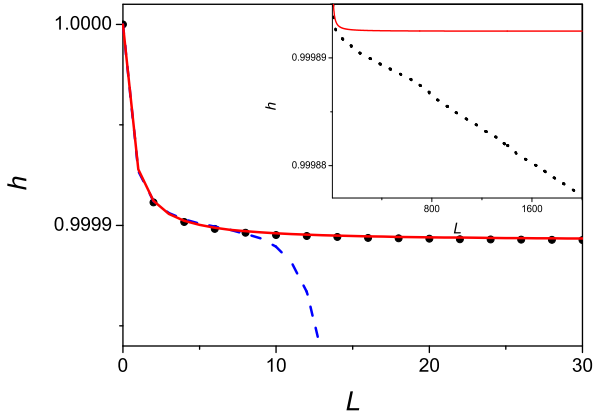}
\caption{The differential entropy vs the length of words. The solid
line is the analytical result, Eq.~(\ref{EntroMain}), for the
correlation function $K(r)=0.01/r^{1.1}$ and $\bar{a}=1/2$, whereas
the dots correspond to the direct evaluations of the same
Eq.~(\ref{EntroMain}) for the numerically constructed sequence (of
the length $M=10^8$ and the cut-off parameter $r_c=10^4$) by means
of conditional probability function~(\ref{Approx CP}) and the
numerically evaluated correlation function $K(r)$ of the constructed
sequence. The dashed line is the differential entropy,
Eqs.~(\ref{entro_block}) and (\ref{entro_diff}), plotted by using
the numerical estimation of probability $P(a_{1},\ldots,a_{L}) $ of
the $L$-blocks occurring in the same  sequence. The inset
demonstrate the linear dependence of differential entropy at large
$L$ governed by fluctuations of the correlation function.}
\label{Weak_cor}
\end{figure}

As an illustration of result (\ref{EntroMain}), in
Fig.~\ref{Weak_cor} we present  the plot of the differential entropy
versus the length $L$. Both numerical and analytical results (the
dotted and solid curves) are presented for the power-law correlation
function $K(r)=0.01/r^{1.1} $. The cut--off parameter $r_c$ of the
power-law function for numerical generation of the sequence,
coinciding with the memory length of the chain, is $10^4$. The good
agreement between the curves is the manifestation of adequateness of
the additive Markov chain approach for studying entropy properties
of random chains. The abrupt deviation of the dashed line from the
upper analytical and numerical curves at $L\sim 10$ is the result of
violation of inequality~(\ref{Strong2}) and a manifestation of
quickly growing errors in the entropy estimation by using the
probability $P(a_{1},\ldots,a_{L}) $ of the $L$ blocks occurring.
Note that violation of Eq.(2) does not depend on the choice of the
model parameter. It only depends on the length $M$ of the random
sequence.

In the main panel of Fig.~\ref{Weak_cor} the deviation of numerical
curve from analytical one is nearly absent. Nevertheless in the
large scale, presented in the inset, a systematic linear deviation
of numerical result from the analytical one is clearly seen.
Explication of this phenomenon is given in the next section while
discussing finite random sequences.

\begin{figure}[h!]
\includegraphics[width=0.5\textwidth]{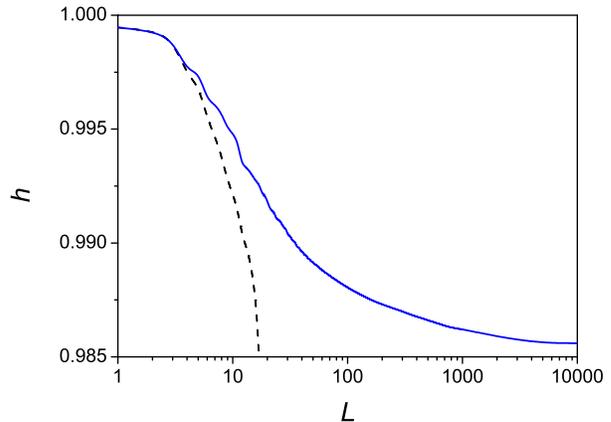}
\caption{Differential entropy $h$ vs length $L$ for R3 chromosome of
\emph{Drosophila melanogaster} DNA of length $M\simeq2.7\times
10^7$. The solid line is obtained by using Eq.~(\ref{EntroMain})
with numerically evaluated correlation function
Eq.~(\ref{def_cor1}). The dashed line is the differential entropy,
Eqs.~(\ref{entro_block}) and (\ref{entro_diff}), plotted by using
the numerical estimation of probability $P(a_{1},\ldots,a_{L})$ of
the $L$ blocks occurring in the same  sequence. } \label{Droso}
\end{figure}

Our next illustration of applicability of the developed theory deals
with the DNA sequence of the R3 chromosome of \emph{Drosophila
melanogaster}. In Fig.~\ref{Droso} the plot of the differential
entropy versus the length $L$ is presented. We see that coincidence
of the two approaches only holds for $L\lesssim 5-6$ units. It is
difficult to do a single-valued conclusion of which factor,
finiteness of the chain and violation of Eq.~(\ref{Strong2}) or
strength of correlations, is more important for discrepancy between
two theories. Nevertheless, even observed coincidence between two
curves seems rather astonishing.

Markov's chains with step-wise memory functions and a larger class
of \emph{permutable} chains are invariant under \emph{decimation}
procedure~\cite{UYa}. Chains whose conditional probability functions
are independent of the order of symbols in the $N$ word preceding a
generated symbol are referred to as permutable. The decimation is a
reduction of a random sequence by regular or random removing some
part of symbols from the whole chain. It was shown in Refs.
\cite{UYa,highord} that after decimation the correlation function of
indicated classes of sequences is invariant up to the new reduced
memory length $N^*=\lambda N$, where $\lambda$ is the relative non
removed part of symbols in the chain. Hence, after decimation
Eq.~(\ref{Entro_Markov2}) does not change its form, but instead of
$N$ we have only to put the new memory length $N^*$.

\section{Finite random sequences}

The relative average number of unities $\bar{a}$, correlation
functions and other statistical characteristics of random sequences
are deterministic quantities only in the limit of their infinite
lengths. It is the direct consequence of the law of large numbers.
If the length $M$ of the sequence is finite, the set of numbers
$a_1^M$ cannot be considered anymore as ergodic sequence. In order
to restore its status we have to introduce an \emph{ensemble} of
finite sequences $\{a_1^M\}_p, p \in \mathbb{N} =0,1,2,...$.
However, we would like to retain the right to examine \emph{finite}
sequences even if approximately by using a single finite chain. So,
for a finite chain we have to replace definition~(\ref{def_cor1}) of
the correlation function by the following one,
\begin{eqnarray} \label{CorFin}
&&C_M(r)=\frac{1}{M-r}
\sum_{i=0}^{M-r-1}(a_i-\bar{a})(a_{i+r}-\bar{a}), \\
&&\bar{a}=\frac{1}{M}\sum_{i=0}^{M-1}a_i. \nonumber
\end{eqnarray}

Now the correlation functions and $\bar{a}$ are random quantities
depending on a particular realization of the sequence $a_1^M$. Their
fluctuations can contribute to the entropy of finite random chains
even if the correlations in the random sequence are absent. It is
well known that the order of relative fluctuations of additive
random quantity (as, e.g. the correlation function
Eq.~(\ref{CorFin})) is $1/\sqrt{M}$.

Below we give more rigorous justification of this explanation and
show its applicability to our case. Let us present the correlation
function $C_M(r)$ as the sum of two components,
\begin{equation}\label{CorrelSquar}
C_M(r)= C(r)+C_{f}(r),
\end{equation}
where the first summand $C(r)=\lim_{M\rightarrow\infty} C_M(r)$ is
the correlation function determined by Eqs.~(\ref {def_cor1}) and
(\ref {CorFin}), obtained by averaging over the sequence with
respect to the index $i$, numbering the elements $a_{i}$ of the
sequence $\mathbb{A}$; and the second one, $C_{f}(r)$, is a
fluctuation--dependent contribution. The function $C(r)$ can be also
presented as the ensemble average $C(r)=\langle C_M(r) \rangle$ due
to ergodicity of the sequence.

Now we can find a connection between variances of $C_M(r)$ and
$C_{f}(r)$. Taking into account that the correlations are weak and
neglecting their contribution into $C_{f}(r)$  we have
\begin{equation}\label{CorrelSquar1}
\langle C^2_{M}(r) \rangle = C^2(r) + \langle C^2_{f}(r) \rangle.
\end{equation}

In order to obtain the last equation we used Eq.~(\ref
{CorrelSquar}) and the property of the function $\langle C_f(r)
\rangle =0$ at $r\neq 0$. The mean fluctuation of the squared
correlation function $C^2_f(r)$ is
\begin{eqnarray} \label{Fluct}
&&\langle C^2_f(r)\rangle =
\frac{1}{(M-r)^2}\sum_{n,m=0}^{M-r-1}\langle(a_n-\bar{a})(a_{n+r}-\bar{a}) \nonumber\\
&&\times(a_m-\bar{a})(a_{m+r}-\bar{a})\rangle.
\end{eqnarray}

Neglecting correlations between the elements $a_n$ and taking into
account that only the terms with $n=m$ give nonzero contribution
into the result we easily obtain
\begin{equation}\label{FluctFin}
\langle K^2_f(r) \rangle = \frac{\langle C^2_f(r)\rangle}{C^2_f(0)},
\,\, \langle K^2_f(r) \rangle = \frac{1}{M-r}\simeq \frac{1}{M}.
\end{equation}

Note that Eq.~(\ref{FluctFin}) is obtained by means of averaging
over the ensemble of chains. This is the shortest way to obtain the
desired result. At the same time, for numerical simulations we used
only averaging over the chain as is seen from Eq.~(\ref{CorFin}),
where summation over the sites $i$ of the chain plays the role of
averaging.

Note also that different symbols $a_i$ in Eq.~(\ref{Fluct}) are
correlated. It is possible to show that contribution of their
correlations to $\langle K^2_f(r) \rangle$ is of order $R_{c}/M^2\ll
1/M$.

The fluctuating part of entropy, proportional to $\sum_{r=1}^L
K^2_f(r)$, should be subtracted from Eq.~(\ref{EntroMain}), which is
only valid  for the infinite chain.
Thus, Eqs.~(\ref{CorrelSquar1}) and~(\ref{FluctFin}) yield the
differential entropy of the \emph{finite} binary weakly correlated
random sequences
\begin{equation}\label{EntroFin}
h_L=  h_0 - \frac{1}{2\ln2}\left[\sum_{r=1}^L K_M ^2(r)-\log_2
\frac{M}{M-L}\right].
\end{equation}

It is clear that in the limit $M\rightarrow\infty$ this function
transforms into Eq.~(\ref{EntroMain}).  When $L\ll M$ the last term
in Eq.~(\ref{EntroFin}) takes the form $L/M$ and describes the
linear decreasing entropy in the inset of Fig.~\ref{Weak_cor}.

The squared correlation function $K^2_M(r)$ is usually a decreasing
function of $r$, whereas the function $K^2_f(r)$ is an increasing
one. Hence, the terms $\sum_{r=1}^L K_M ^2(r)$ and $\log_2 [M/(M-L]$
being concave and convex functions describe competitive
contributions to the entropy. It is not possible to analyze all
particular cases of their relationship. Therefore we indicate here
the most interesting ones keeping in mind a monotonically decreasing
correlation function. An example of such type of function,
$K(r)=a/r^{b}, \, a>0, \,b \geqslant 1 $, was considered above.

If the correlations are extremely small and compared with the
inverse length $M$ of the sequence, $K_M ^2(1) \sim 1/M$, the
fluctuating part of entropy exceeds the correlation one nearly for
all values of $L>1$.
\begin{figure}[h!]
\includegraphics[width=0.5\textwidth]{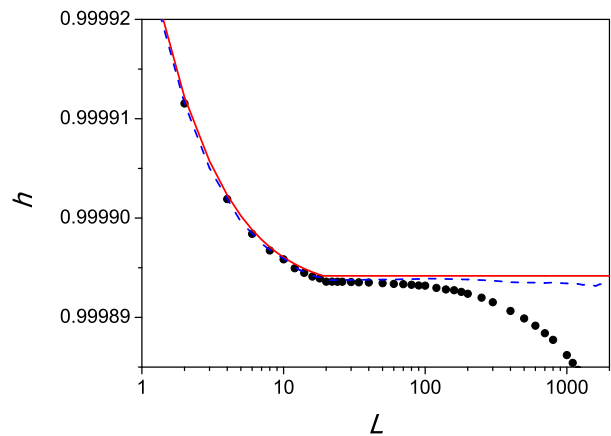}
\caption{The differential entropy vs the length of words.
 The solid line is the analytical result for the
correlation function $K(r)=0.01/r^{1.1} $, whereas the dots
correspond to the direct numerical evaluations Eq.~(\ref{EntroMain})
for the numerically constructed sequence of the length $M=10^8$ and
the cut-off parameter $r_c=20$. The dashed line is the differential
entropy with fluctuation correction described by
Eq.~(\ref{EntroFin}).  } \label{Short Mem}
\end{figure}

With increasing of $M$ (or correlations), when the inequality $K_M
^2(1)> 1/M$ is fulfilled, there is at list one point where the
contribution of fluctuation and correlation parts of entropy are
equal. For monotonically decreasing function $K(r)$ there is only
one such point. Comparing the functions in square brackets in
Eqs.~(\ref{EntroFin}) we find that they are equal at some $L =
R_{s}$, which hereafter will be referred to as a stationarity
length. If $L \ll R_s$, the fluctuations of the correlation function
are negligibly small with respect to its magnitude, hence the finite
sequence may be considered as quasi-stationary. At $ L \sim R_{s}$
the fluctuations are of the same order as the genuine correlation
function $K^2(r)$. Here we have to take into account the fluctuation
correction due to finiteness of the random chain. At $ L > R_{s}$
the fluctuating contribution exceeds the correlation one.

The other important parameter of the random sequence is the memory
length $N$. If the length $N$ is less than $R_{s}$, we have no
difficulties to calculate the entropy of finite sequence, which can
be considered as quasi-stationary. This case is illustrated in
Fig.~\ref{Short Mem}. If the memory length exceeds the stationarity
length $R_{s} \lesssim N$, we have to take into account the
fluctuation correction to the entropy.
\begin{figure}[h!]
\includegraphics[width=0.5\textwidth]{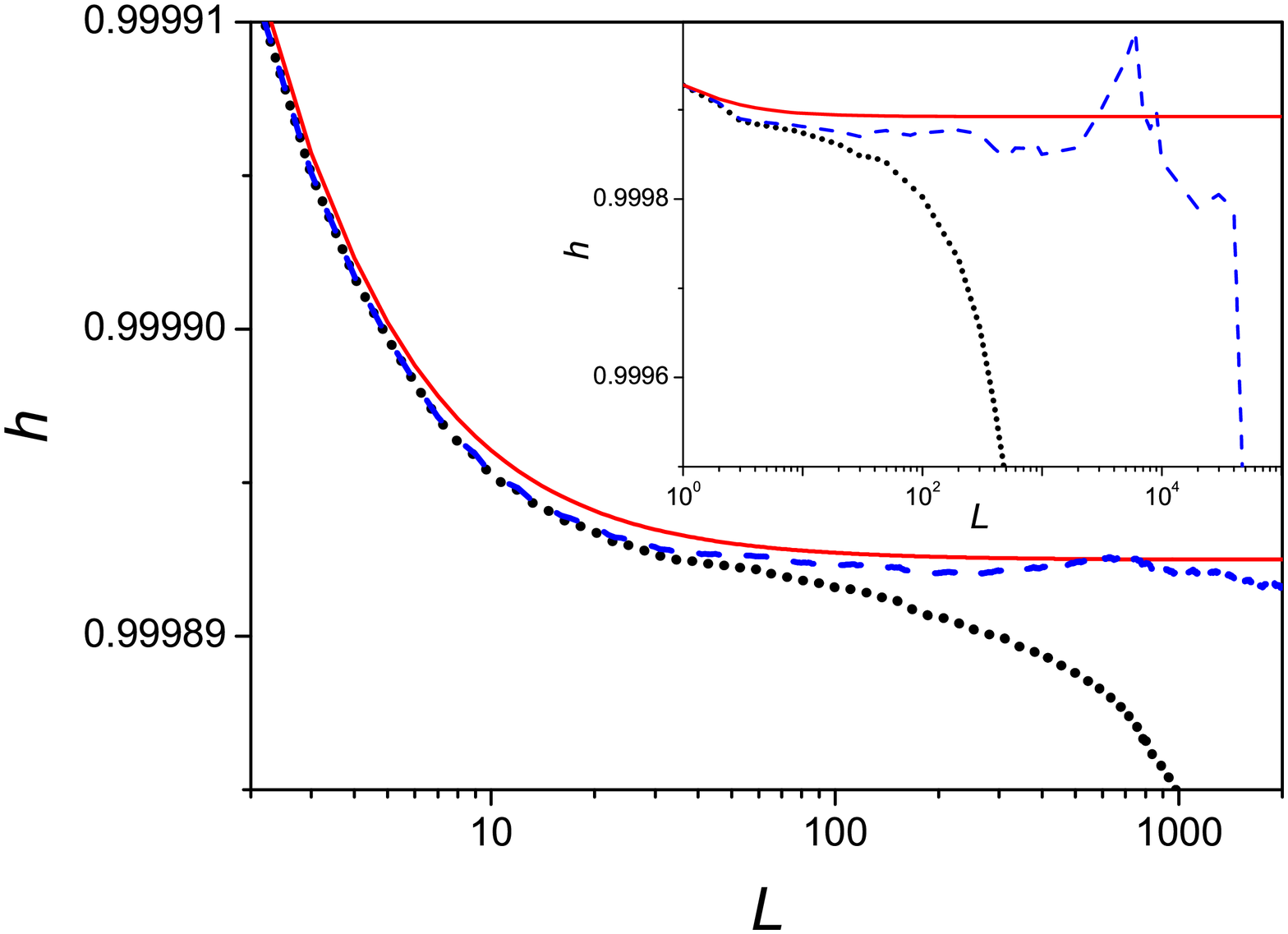}
\caption{The differential entropy vs the length of words in $L$-axis
log scale. The solid and dotted curves are the same as in the main
panel of Fig.~\ref{Weak_cor}. The dashed line corresponds to the
direct evaluations of Eq.~(\ref{EntroMain}) for the sequence
numerically constructed  by means of Eq.~(\ref{Approx CP}) with
fluctuation correction (\ref{EntroFin}) and the cut-off parameter
$r_c=10^4$. The inset demonstrates the large $L$ region for the
sequence of the length $M=10^6$.} \label{pow-long}
\end{figure}

In Fig.~\ref{pow-long} the plot of the differential entropy versus
the length of words is shown as an illustration of importance of
this correction. Both numerical and analytical results are presented
for the same power-law correlation function as in
Figs.~\ref{Weak_cor} and \ref{Short Mem}. Comparing sums of squared
correlation function $K(r)=0.01/r^{1.1}$ with
contribution~(\ref{Fluct}), proportional to $\log_2 [M/(M-L]$, we
find that they are equal at $R_s\approx 10^4$. A graphical
confirmation of this result is shown in the inset of
Fig.~\ref{pow-long}. We can conclude that the dashed lines better
approximate theoretical solid curves than dotted lines (till
$L\approx10^4$). The nonmonotone decrease of $h(L)$ is due to the
fluctuation of random quantity, the entropy of the finite sequence.

\section{Application to written texts}

A theory of additive Markov chains with long-range memory was used
for a description of correlation properties of literary
texts~\cite{MUYaG}. The coarse-grained naturally written texts were
shown to be \emph{strongly} correlated sequences that possess
anti-persistent properties at small distances (in the region $L\leq
300$ of grammatical rules action).  At long distances (in the region
$L\geq 300$ of semantic rules action) they  manifest \emph{weak}
persistent power-law correlations. It is clear that our model of the
additive Markov chain can only claim to describe the weak power-law
part of entropy, proportional to $L^{-\gamma}$.

Ebeling and Nicolis~\cite{EbelNic} and Sch\"{u}rmann and
Grassberger~\cite{Schur} suggested the empirical form of entropy for
written texts
\begin{equation}\label{Entro_Markov3}
h_L=h+c \frac{\log_2 L}{L^{\gamma}},  \quad \gamma > 0.
 \end{equation}

There emerges a natural question about the origin of this
dependence. The partial answer to the question is as follows.  The
entropy of the Markov chain with step-wise memory function
~(\ref{prob step}) in the limit of \emph{ strong } correlations,
$N\ln N(1-2\mu)\ll 4\mu$, was obtained in Ref.~\cite{DMBUYa}
\begin{equation} \label{strong_pers}
h_L=  h +  c \frac{\log_2 L}{L}.
\end{equation}

After comparing the results of Eqs.~(\ref{Entro_Markov2})
and~(\ref{strong_pers}) with that of Eq.~(\ref{Entro_Markov3}), it
becomes clear that the term $\log_2 L$ describes strong short-range
correlations and the power-law term $L^{-\gamma}$ is responsible for
weak long-range correlations. So, we need a combined model that
could unify two approaches of the additive Markov chain exposed
above and the Markov chain with a step-wise memory function.

The answer to the question of which part of the correlation or
memory function is responsible for the decimation invariance is
still unsolvable.

\section{Conclusion and perspectives}

(i) The main result of the paper, the differential entropy of the
stationary ergodic binary weakly correlated random sequence
$\mathbb{A}$ is given by Eq.~(\ref{EntroMain}). The other important
point of the work is the calculation of the fluctuation contribution
to the entropy due to finiteness of random chains, the last term in
Eq.~(\ref{EntroFin}).

(ii) In order to obtain Eq.~(\ref{EntroMain}) we used an assumption
that the random sequence of symbols is the Markov chain.
Nevertheless, the final result contains only the correlation
function, does not contain the conditional probability function of
the Markov chain. This allows us to suppose that
result~(\ref{EntroMain}) and the region of its applicability is
wider than the assumptions under which it is obtained~\cite{Apost}.

(iii) To obtain Eq.~(\ref{EntroMain}) we have supposed that
correlations in the random chain are weak. This is not a very severe
restriction. Many examples of such systems, described by means of
the pair correlator are given in Ref.~\cite{IzrKrMak}. The randomly
chosen example of DNA sequences supports this conclusion. The
strongly correlated systems, which are opposed to weakly correlated
chains, are nearly deterministic. For their description we need
completely different approach. Their study is beyond the scope of
this paper.

(iv) The developed theory opens a way for constructing a more
consistent and sophisticated approach describing the systems with
long-range correlations. Namely, Eq.~(\ref{EntroMain}) can be
considered as expansion of the entropy in series with respect to the
small parameter $\delta$, where the entropy $h_0$ of the
non-correlated sequence is the zero approximation. Alternatively,
for the zero approximation we can use the exactly solvable model of
the $N$-step Markov chain with the conditional probability function
of words occurring taken in the form of the step-wise function,
Eq.~(\ref{prob step}). Another way to choose the zero approximation
can be based on CPF obtained from probability of the bloc occurring
Eq.~(\ref{entro_block}).

(v) In this paper we have considered the random sequences with the
binary space of states, but almost all results can be generalized to
non-binary sequences and can be applied for describing natural
written and DNA texts.

(vi) Our consideration can be generalized to the Markov chain with
the infinite memory length $N$. In this case we have to impose a
condition on the decreasing rate of the correlation function and the
conditional probability function at $N\rightarrow \infty$.

\begin{acknowledgments} We are grateful for the very helpful and fruitful
discussions with  A.~A.~ Krokhin, G.~M.~ Pritula, S.~V.~Denisov,
S.~S.~ Apostolov, and Z.~A.~ Mayzelis.
\end{acknowledgments}

\appendix

\section{ }

Here we prove Eq.~(\ref{p_i(L)}) using Eq.~(\ref{Approx CP}) as a
starting point. It follows from definition (\ref{soglas}) of the
conditional probability function
\begin{equation}\label{App1}
P(1|a_{i-N+1}^{i-1})=\frac{P(a_{i-N+1}^{i-1},1)} {
P(a_{i-N+1}^{i-1})}.
\end{equation}
Adding  the symbol $a_{i-N}$ to the string $(a_{i-N+1}^{i-1},1)$ we
have
\begin{equation}\label{App2}
P(1|a_{i-N+1}^{i-1})=\frac{P(0,a_{i-N+1}^{i-1},1)+P(1,a_{i-N+1}^{i-1},1)}
{ P(a_{i-N+1}^{i-1})}.
\end{equation}
Replacing here the probabilities $P(a_{i-N},a_{i-N+1}^{i-1},1)$ with
the CPF $P(1|a_{i-N},a_{i-N+1}^{i-1})$ from equation similar to that
of Eq.~(\ref{App1}),
\begin{equation}\label{App2'}
P(1|b,a_{i-N+1}^{i-1})=\frac{P(b,a_{i-N+1}^{i-1},1)} {
P(b,a_{i-N+1}^{i-1})},\,\,b = (0,1),
\end{equation}
after some algebraic manipulations, we get
%
%
%
\begin{eqnarray} \label{App3}
&&P(1|a_{i-N+1}^{i-1})= \bar{a} + \sum_{r=1}^{N-1} F(r)(a_{i-r}-
\bar{a}) \\
&&+\frac{F(N)}{P(a_{i-N+1}^{i-1})}\left[(1-\overline{a})
 P(1,a_{i-N+1}^{i-1}) - \overline{a} P(0,a_{i-N+1}^{i-1})\right].\nonumber
\end{eqnarray}
%

From the compatibility condition for the Chapman-Kolmogorov equation
(see, for example, Ref.~\cite{gar}),
\begin{equation}\label{App4}
P(a_{i-N+1}^{i})=\sum_{a_{i-N}=0,1}P(a_{i-N}^{i-1}),
P_{N}(a_i|a_{i-N}^{i-1}),
\end{equation}
it follows that its solution is given by
\begin{equation}\label{App5}
P(k)=\overline{a}^k(1-\overline{a})^{N-k}+O(\delta).
\end{equation}
Here $P(k)$ is the probability to have $k$ units and $(N-k)$ zeros
at fixed sites of the $N$-word. Therefore, the last term in the
square brackets of Eq.~(\ref{App3}) vanishes in the main
approximation, so that the difference $\left[(1-\overline{a})
 P(1,a_{i-N+1}^{i-1}) - \overline{a} P(0,a_{i-N+1}^{i-1})\right]$ is
of order of $\delta$. Hence, we have to neglect the third term in
the right-hand side of Eq.~(\ref{App3}) because it is of the second
order in $\delta$. So, Eq.~(\ref{p_i(L)}) is proven for $L=N-1$. By
induction the equation can be written for arbitrary $L$.


\end{document}